\documentstyle[aaspp4]{article} 
\def \sun {$_{\scriptscriptstyle\odot}$} 

\lefthead{FRYER \& KALOGERA} 
\righthead{Black Hole Masses} 
\begin{document} 
\begin{center} 
Submitted to {\em The Astrophysical Journal}
\end{center}
\vspace{1.cm}

\title{Theoretical Black Hole Mass Distributions} 
\author{Chris L. Fryer} 
\affil{Lick Observatory, University of California
Observatories, \\ Santa Cruz, CA 95064 \\ cfryer@ucolick.org}
\authoremail{cfryer@ucolick.org} 

\author{Vassiliki Kalogera}
\affil{Harvard-Smithsonian Center for Astrophysics, 60 Garden
St., Cambridge, MA 02138 \\ vkalogera@cfa.harvard.edu}
\authoremail{vkalogera@cfa.harvard.edu}

\begin{abstract}

We derive the theoretical distribution function of black hole masses 
by studying the formation processes of black holes.  We use the 
results of recent 2D simulations of core-collapse to obtain 
the relation between remnant and progenitor masses and fold it with 
an initial mass function for the progenitors.  We examine how the
calculated black-hole mass distributions are modified by (i) strong wind
mass loss at different evolutionary stages of the progenitors, and (ii)
the presence of close binary companions to the black-hole progenitors. 
Thus, we are able to derive the binary black hole mass distribution.
The compact remnant distribution is dominated by neutron stars in the 
mass range 1.2--1.6\,M\sun\, and falls off exponentially at higher remnant
masses.  Our results are most sensitive to mass loss from winds which is 
even more important in close binaries.  Wind mass-loss causes the black hole 
distribution to become flatter and limits the maximum possible black-hole 
mass ($\lesssim 10-15$\,M\sun).  We also study the effects of the 
uncertainties in the explosion and unbinding energies for different 
progenitors.   The distributions are continuous and extend over a broad 
range.  We find no evidence for a gap at low values ($3-5$\,M\sun) or for a 
peak at higher values ($\sim 7$\,M\sun) of black hole masses, but we argue 
that our black hole mass distribution for binaries is consistent with the 
current sample of measured black-hole masses in X-ray transients.  We discuss 
possible biases against the detection or formation of X-ray transients 
with low-mass black holes.  We also comment on the possibility
of black-hole kicks and their effect on binaries. 

\end{abstract}


\section{Introduction}

Soft X-ray transients provide a window into the nature of compact objects,
and in particular, stellar mass black holes.  During the quiescent state
of emission (low X-ray luminosity), optical/infrared observations of the 
compact object's companion allow the measurement of the mass function and 
other binary parameters. In the case of small companion masses (late 
spectral type) the value of the mass function, $f(M_{\rm X})$, sets a 
strong lower limit to the mass of the compact object: 
 \begin{equation}
 M_{\rm X}~=~f(M_{\rm X})\,\frac{(1+q)^2}{ {\rm sin}^3i}~ > ~f(M_{\rm X}),
 \end{equation}
 where $q\equiv M_c/M_{\rm X}$ is the mass ratio, and $i$ is the
inclination of the binary orbit. For several systems, this lower limit
exceeds 3\,M\sun, the optimum (independent of the equation of state of
matter at densities exceeding nuclear densities)  upper limit to the
gravitational mass of a neutron star (NS) (\cite{R74} 1974;
\cite{KB96}1996), strongly suggesting that these compact objects are black
holes (for a recent review see \cite{C98} 1998). Indeed, at present, there
is little doubt that stellar mass black holes (BH)  exist, and that they
play an important role in high-energy astrophysics (e.g., X-ray binaries,
gamma-ray bursts).

Currently there are 9 X-ray transients thought to contain a BH accreting
matter from a low mass companion. They are listed in Table 1 along with
the range of BH mass estimates reported in the literature. Most of the
data have been taken from \cite{B98}(1998) who presented a detailed
statistical study of the BH mass measurements available at the time and
the associated errors (dominated by uncertainties in the inclination
angle). Recently, revised measurements of the mass function for GRO
J1655-40 (\cite{Phi99}1999) and GRO J0422+32 (\cite{Har99}1999) have led
to a decrease in the lower mass limits. In Table 1 we also include two new
systems 4U 1543-47 (\cite{O98}1999) and GRS 1009-45 (\cite{Fil99}1999),
although their mass estimates are still rather uncertain.

\cite{B98}(1998) studied the statistical properties of the early sample of
measured BH masses (excluding 4U 1543-47 and GRS 1009-45, and the revised
masses for GS1124-68 and GRO J0422+32) and concluded that almost all of
the estimates are consistent with a narrow range of BH masses around
7\,M\sun\, and that a gap between 3 and 5\,M\sun\, is present in the
distribution. In this paper we calculate the expected BH mass
distribution and its dependence on a number of physical elements (stellar
winds and binary evolution) based on the current theoretical understanding
of BH formation.

We consider BH formation during the collapse of massive stars. This
process can proceed in two different ways: either the massive star
collapses directly into a BH without a supernova explosion, or an explosion
occurs, but its energy is too low to completely unbind the stellar envelope, 
and a significant part of the star falls back onto the proto-neutron star
forming a BH (\cite{C71}1971; \cite{W88}1988; {\cite{F99a}1999a). An
alternative path involving the collapse of a NS into a BH through
hypercritical accretion (e.g., in a common envelope or in binary mergers)
has often been discussed in recent years (see \cite{FWH99}1999 for a
review). However, even under the assumption that BH formation occurs every
time a NS goes through a common-envelope phase or a merger, the event
rates for NS accretion-induced collapse are lower than those for massive
star collapse events by factors of 10--100 (\cite{FWH99}1999). Therefore,
we assume that BH formation is dominated by stellar collapse.

In this paper we use the results of up-to-date 2D hydrodynamic simulations
of core collapse (\cite{F99a}1999a) for explosion energies and remnant
masses to obtain theoretically expected BH mass distributions. In \S\,2, 
we outline the steps involved in constructing such distributions and 
discuss their associated uncertainties. We present our 
results in \S\,3 and conclude that the BH mass distribution and particularly 
its slope and upper cut-off is most sensitive to the effects of stellar 
winds in binaries.  No evidence for a gap is found at low BH masses.  
In the last section we compare our results to observations and discuss
some possible biases and uncertainties in the observed sample.

\section{Black Hole Mass Distributions}

\subsection{Outline of the Calculation}

The calculation of theoretically expected BH mass distributions involves a
sequence of steps. We derive the remnant mass as a function of progenitor
mass\footnote{Since BH progenitors can lose mass as they evolve because of
winds and binary mass transfer, the term ``progenitor mass'' is not
uniquely defined.  To avoid confusion, we use progenitor mass to refer to
the initial mass of the BH progenitor unless we explicitly state
otherwise (e.g. progenitor mass {\it at collapse}.)} based on an
energy-budget argument regarding the {\em necessary} and {\em available}
energies to unbind the stellar envelope. We scale the available energy to
the explosion energy calculated from core-collapse simulations. We then
convolve the resulting remnant-progenitor mass relation with the initial
mass function of the progenitors to obtain the BH mass distribution. Each
of these steps, and hence the results, can be affected quantitatively by
the evolution of the progenitor prior to collapse. Before we present our
results and the associated uncertainties, we detail the steps of our
calculation.

\noindent
 {\em Explosion Energy.} To estimate the energy available to unbind the
stellar envelope and its dependence on progenitor properties (mass and
density structure), we are guided by recent 2D core-collapse simulations
(\cite{F99a}1999a), which provide us with the explosion energy as a
function of the progenitor mass.  This explosion energy is determined by
the change in total energy before collapse and at the end of the $\sim$1s
simulation of the material that is initially part of the exploding shock
(for most simulations, this is the entire star beyond the 1.2\,M\sun\,
proto-neutron star).  Mass loss (from winds or binary mass transfer) will
not change the explosion energies unless the mass-loss affects the inner
$\sim$3\,M\sun\, of the stellar core.  Hence, the explosion energies as a
function of progenitor mass from \cite{F99a}(1999a) are valid for both
helium and hydrogen stars.  For stars in which mass loss has affected the
core, C/O cores at collapse have been calculated (\cite{LH95} 1995) and we
use these masses to estimate the explosion energy. We find that mass loss
affects the cores of stars only for a small fraction of the BH progenitors
(more massive than $\sim 40$\,M\sun).  For example, a 40\,M\sun\, star with
winds from \cite{LH95}(1995) will have an explosion energy closer to that 
of a 30\,M\sun\, progenitor star without winds.  Such cases are discussed 
in more detail in \S\,2.2.3. 

 \begin{figure}[ht]
 \epsscale{0.8}
 \plotone{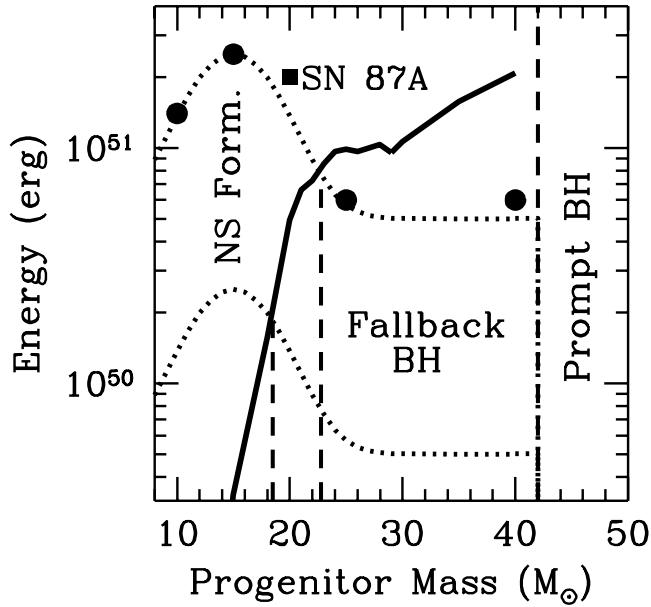}
 \caption{Explosion energy and binding energy as a function of
hydrogen-rich progenitor mass. The circles are explosion energies
calculated from core-collapse simulations (\cite{F99a}1999a), while the
square is the explosion energy inferred for SN 1987A (\cite{W88}1988).  
The upper dotted line is a fit to the numerical results for the explosion
energy.  Since only a fraction of the energy is used to unbind the star,
we plot the lower dotted line, for which we assume that only 10\% of the
explosion energy actually goes into unbinding the star.  The solid line is
the binding energy of the entire stellar envelope outside the inner 3
M\sun.  The point at which the energy available to unbind the star crosses
the binding energy marks the limit where BH formation occurs.}
 \end{figure}

 \begin{figure}[ht]
 \epsscale{0.8}
 \plotone{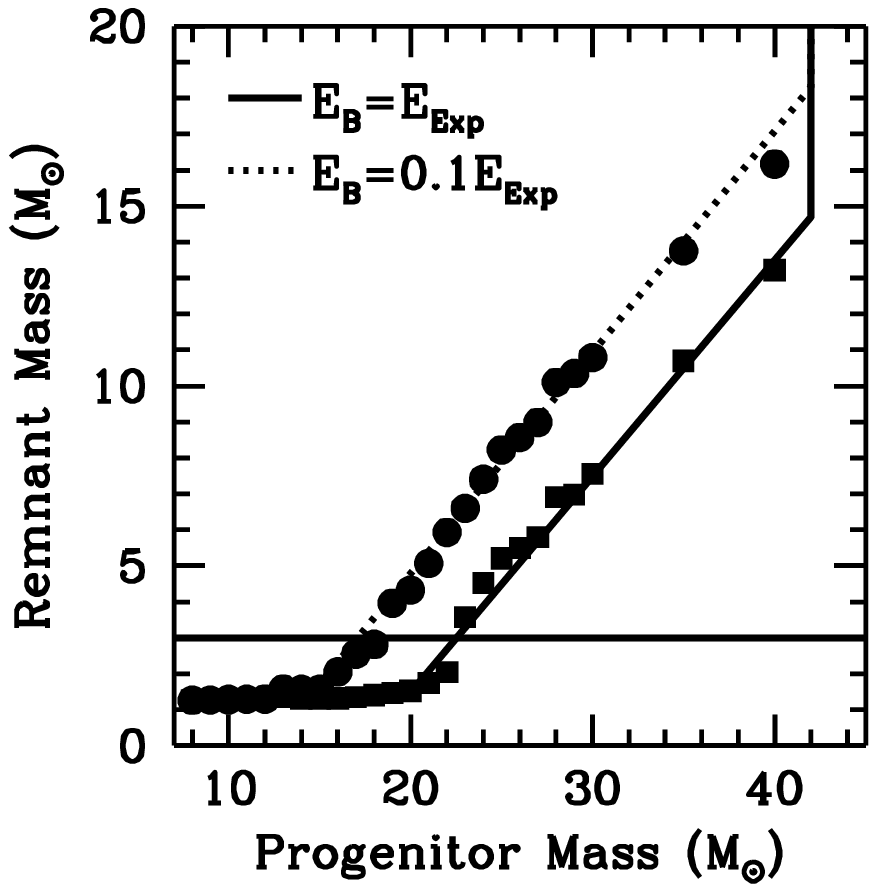}
 \caption{Compact remnant mass as a function of progenitor mass, assuming
the energy, $E_B$, available to unbind the stellar envelope is equal to
the explosion energy (solid line) or equal to 10\% of the explosion energy
(dotted line). For progenitors more massive than 40\,M\sun, there is no
supernova explosion. The star collapses on itself and forms a black hole
with mass equal to its progenitor mass.}
 \end{figure}

Knowing the explosion energy from core-collapse models is not
sufficient to estimate the final compact remnant mass. We also need to
know what fraction of the explosion energy goes into unbinding the star,
and what fraction is transformed into kinetic energy of the ejecta.  In
Figure 1 we plot (dotted lines) two different examples of the energy
available to unbind the star as a function of the progenitor mass. For the
top curve, we assume that 100\% of the collapse energy goes into unbinding
the star and is a fit to the results (black dots)  of \cite{F99a}(1999a).  
The lower curve is a scaled-down version of the top curve assuming that
10\% of the explosion energy is used to unbind the star, the rest being
transformed into kinetic energy of the ejecta.  The simulations show that
above a critical progenitor mass, no supernova occurs, and the entire star
collapses to form a black hole.  A progenitor of $\sim$40\,M\sun\, appears
to be the critical mass for BH formation (see \cite{F99a}1999a for more
details).

\begin{figure}[hp]
 \epsscale{0.70}
 \plotone{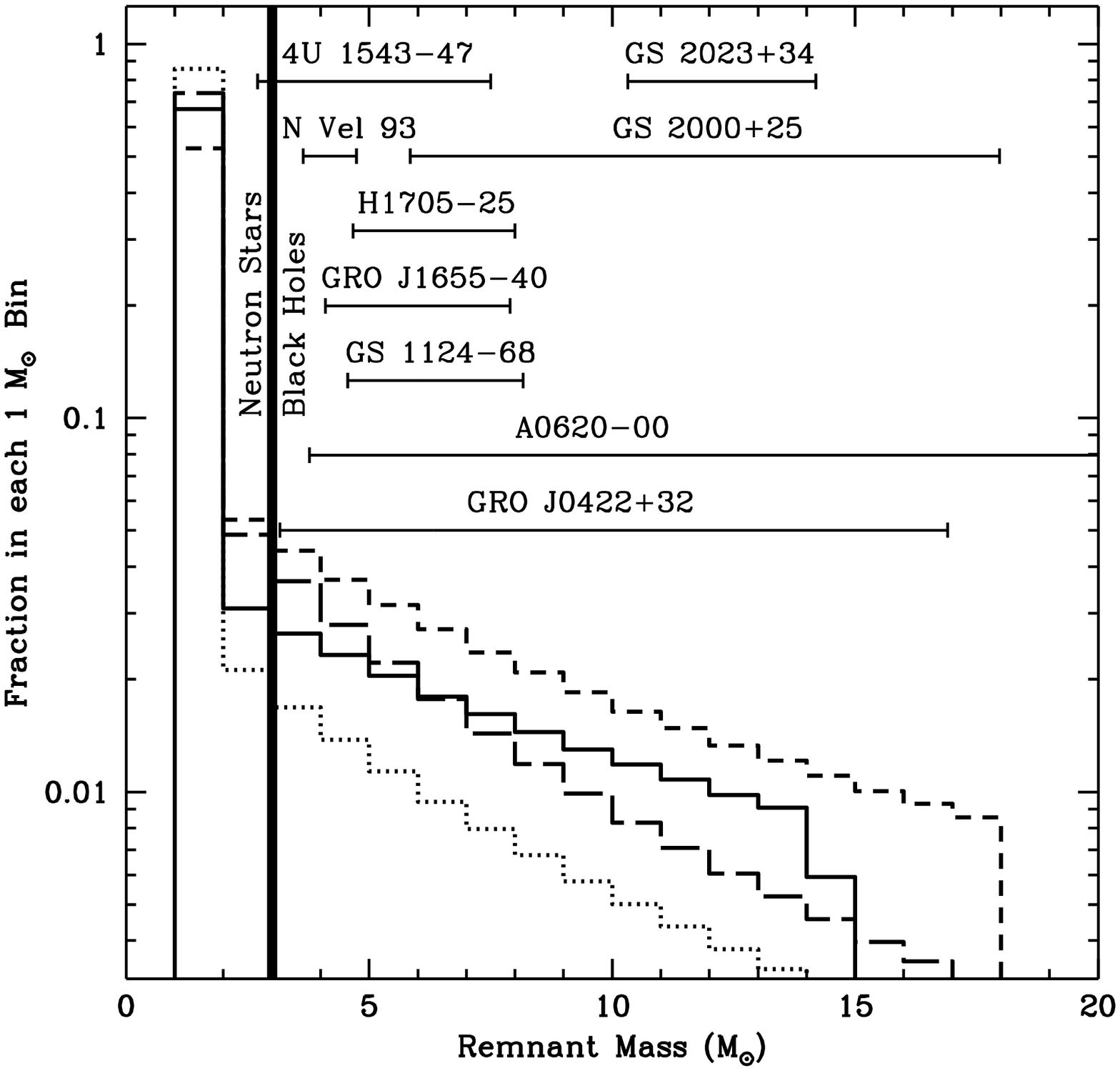}
 \caption{The mass distribution of compact remnants using the energies
from Figure 1, for $f=1$ (solid and dotted lines) and $f=0.1$ (short- and
long-dashed lines). We assume a power law initial mass function with
indices $\gamma=2.0$ (solid and short-dashed lines) and $\gamma=3.0$
(dotted and long-dashed lines). Assuming the maximum NS mass is $\sim
3$M\sun, we find that $\sim 80$\% of compact remnants are neutron stars.}
 \end{figure}

\noindent
 {\em Stellar Binding Energy.}
 The next step is to compare the energy available to eject material to the
actual binding energy of the star, obtained from integrating the stellar
mass profile. For these profiles we use the models of massive stars
calculated by \cite{W95}(1995). In Figure 1 we plot (solid line) the
energy required to unbind all but the inner 3\,M\sun\, for each
progenitor. We assume that the maximum NS mass is 3\,M\sun\, and we thus
calculate an {\em upper limit} to the critical initial progenitor mass
dividing NS from BH formation through fallback. From Figure 1 it is
evident that the critical masses are $\sim$18\,M\sun\, and $\sim
23$\,M\sun\,
for 10\% and 100\% of the collapse energy being available to unbind the
star, respectively.  If the maximum NS mass is lower, as most equations of
state in the literature predict (e.g., \cite{CST94}1994), the binding
energies increase, and the critical progenitor masses decrease.  The
hydrogen envelope accounts for less than $\sim10$\% of this total binding
energy budget. Hence, if a star loses its hydrogen envelope before
collapse but retains its helium core, its binding energy will not change
by more than 10\%.

\noindent
 {\em Remnant mass.}
We can combine the estimated energies available to unbind
the star and the stellar binding-energy profile (binding
energy as a function of mass within the star) to calculate the
remnant mass, $M_{\rm rem}$, as a function of progenitor
mass, $M_{\rm prog}$:
 \begin{equation}
 f \times E_{\rm explosion}~=~\int_{M_{\rm rem}}^{M_{\rm
prog}} E(m)dm, 
 \end{equation}
 where $f$ is the fraction of the total explosion energy used to unbind
the outer layers of the star (down to $M_{\rm rem}$), and $E(m)$ is the
binding energy profile. In this way, we can calculate the final remnant
mass (and the amount of fallback) for a given progenitor mass. For $E(m)$
we use \cite{W95}(1995) massive star models. The remnant masses as a
function of progenitor mass are shown in Figure 2 for $f=0.1$ and $f=1$.  
For very low explosion energies (high masses), most of the star falls back
to form a black hole and the black hole mass is limited by the progenitor
mass at collapse.  In such cases, even if mass loss does not affect the
explosion energy, it dictates the BH mass by limiting the total mass
available to fall back and form a black hole (\S 2.2).

\noindent
 {\em Initial mass function of progenitors.}
 The final step in the calculation of a theoretical BH mass distribution
requires the use of an initial mass function (IMF) for the massive
BH progenitors, which we transform into a remnant mass function using 
the derived remnant-progenitor mass relation (Figure 2):
 \begin{equation}
 F(M_{\rm rem})~=~F(M_{\rm prog})\,\left(\frac{dM_{\rm rem}}{dM_{\rm
prog}}\right)^{-1}. 
 \end{equation}
There is observational evidence that the IMF in the mass ranges of
interest here is very well described by a single power law: $F(M_{\rm
prog}) \propto M_{\rm prog}^{-\gamma}$ (for a Scalo IMF, $\gamma=2.7$).
Recent studies of massive stars claim a range of values for $\gamma$ 
from 1.8 to 3.1 (see \cite{M91}1991) with no evidence for a dependence 
on metallicity (for massive stars).  Because of this wide spread, we 
investigate a range of values for $\gamma$ between 2 and 3 (Figure 3), 
while for most of the calculations we adopt $\gamma=2.7$. The total number 
of BH is quite sensitive to $\gamma$ (to the critical mass dividing NS and 
BH formation as well). Changing its value from 2 to 3 lowers the fraction 
of BH among core-collapse remnants from 30\% to 12\% (see Table 2, rows 1 
and 2).

\subsection{Results and Uncertainties}

\subsubsection{Supernova Energies and Remnant masses}

The core-collapse simulations we use to derive explosion energies follow
just the first $\sim$1s of the explosion (\cite{F99a}1999a). In these
models, the outward moving shock is still within the oxygen layer of the
star at the end of the simulation.  As the shock moves out of the star, a
fraction of its energy is transformed into kinetic energy of the ejecta
and another fraction goes into unbinding the core. Neutrino emission or
absorption and nuclear burning and dissociation also contribute to the
total energy budget but they represent only a minor fraction of it 1\,s
after bounce.  In determining remnant masses we only need an estimate of
the energy fraction, $f$, used to unbind the star. We vary this fraction
between 0.1 and 1 and the total number of BH remnants changes by a factor
smaller than 2 (Table 2, rows 1--4). Recent fallback models by
\cite{MF99}(1999)  suggest that $f$ lies in the range 0.3--0.5. In most of
our models we adopted $f=0.5$.

The binding energy of stars at collapse is also difficult to determine.  
In the rotating stellar models by \cite{HLW99}(1999), the binding energies
(for matter outside the inner 3\,M\sun) are nearly a factor of two lower
than those in the \cite{W95}(1995) non-rotating models The treatment of
convection also introduces uncertainties in the binding energy.  
Currently, we are limited to the grid of pre-collapse stars produced by
\cite{W95}(1995) and by the uncertainties in their assumptions for
convection, rotation, etc.  Lowering the binding energies of all stars by
a factor of 2 is equivalent to raising $f$ by a factor of 2.  In this
manner, varying $f$ gives us an idea of the uncertainties caused by
varying the magnitude of the binding energies.  The trend of increasing
binding energy with increasing mass is secure, and it is this trend that
plays the most crucial role in our determination of the BH mass
distribution determination.

\begin{figure}[ht]
 \epsscale{0.80}
 \plotone{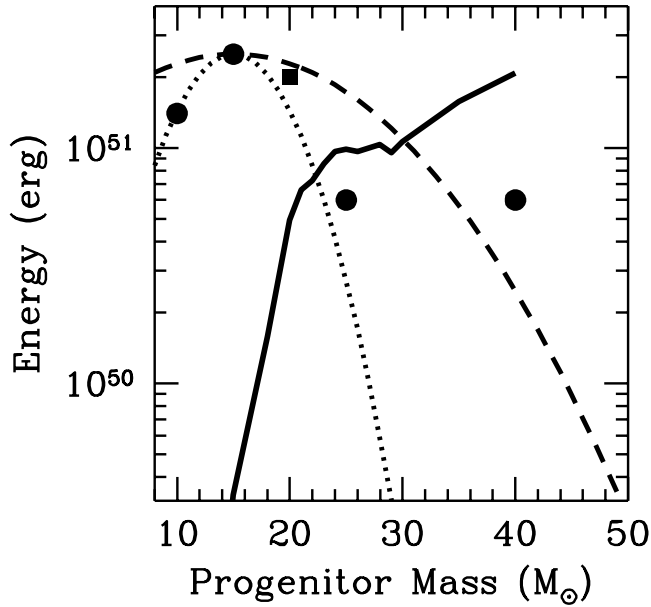}
 \caption{Explosion energy as a function of progenitor mass. The energy is
assumed to rise and fall off faster (dotted line) and slower (dashed line)
than our standard case (Figure 1, top curve).  We plot the results from
core-collapse simulations (circles) and SN 1987A (square) and the binding
energy as a function of progenitor mass (solid line) for comparison.}
 \end{figure}

The accuracy of the determined explosion energies depends on the current
understanding of the supernova mechanism. Progress in this field indicates
that the mechanism is sensitive to detailed physical processes and
ingredients such as the equation of state, neutrino cross sections and
transport, general relativity, and multi-dimensional effects (see
\cite{Bur98}1998 and \cite{M99}1999 for reviews). Until these are worked
out, no ``final'' answer to the question of the explosion energy as a
function of progenitor mass can be given.  In fact, it is likely that for
a given progenitor star mass, there is a range of possible supernova
energies depending on the rotation of the progenitor (\cite{FH99}1999).  
As with the binding energies, we can understand the effect of raising or
lowering the explosion energy by varying $f$.  More importantly, the
trend of an increasing core-collapse energy up to a progenitor mass
$\sim$15\,M\sun, followed by a decrease for more massive progenitors is
expected to remain unchanged. To assess the effect of such uncertainties
on the BH mass distribution, we vary the slope of this energy rise and
decline (only the rate of decline is important for determining BH masses)
with progenitor mass (Figure 4). We consider two cases, one of fast
(dotted line) and one of slow (dashed line) decay compared to our standard
case (Figure 1, top curve). The total number of black holes is not
dramatically sensitive to the choice of this slope (Table 2, e.g., rows 
5,11,12). However, the case of fast decline produces many more black holes
with masses in excess of 10\,M\sun\, (Table 2, Figs.\ 5, 6).  For this
simple case, the critical progenitor mass separating fallback and prompt
BH formation decreases to $\sim$28\,M\sun\, (Figure 5) from
$\sim$40\,M\sun\, for our standard case.  We note that once we include the
effects of stellar winds and close binary evolution (see \S\,2.2.2 and
2.2.3), the differences caused by varying the dependence of supernova
energy on progenitor mass drops to the 20--30\% level for the fraction of
BH remnants in most mass ranges (Table 2).  We ignore any mass loss that 
might occur after black hole formation (e.g. \cite{MF99}) because, at 
this time, we are unable to determine how often (if at all) such explosions 
occur and do not know the amount of mass ejected even if an explosion occurs.

\begin{figure}[ht]
 \epsscale{0.80}
 \plotone{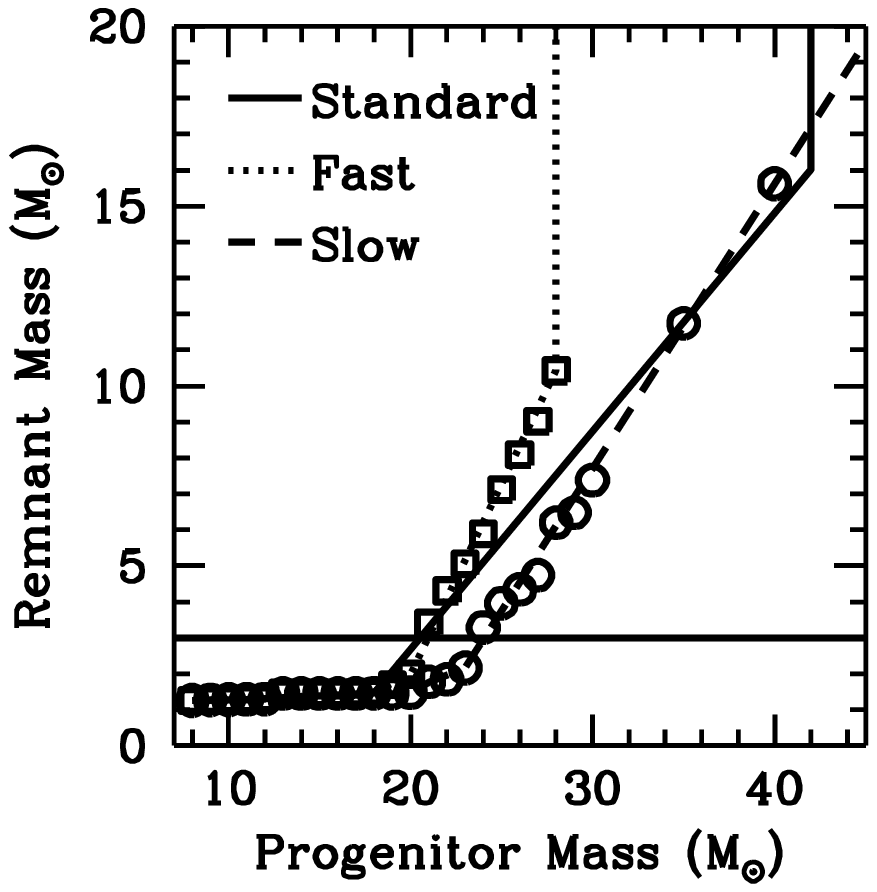}
 \caption{Remnant mass as a function of progenitor mass assuming that 50\%
of the explosion energy from Figures 1 and 4 is used to unbind the star.
Three different cases for the slopes of rise and decline of the explosion
energy are shown: standard (solid line), fast (dotted line) and slow
(dashed line). }
 \end{figure}

\begin{figure}[ht]
 \epsscale{0.70}
 \plotone{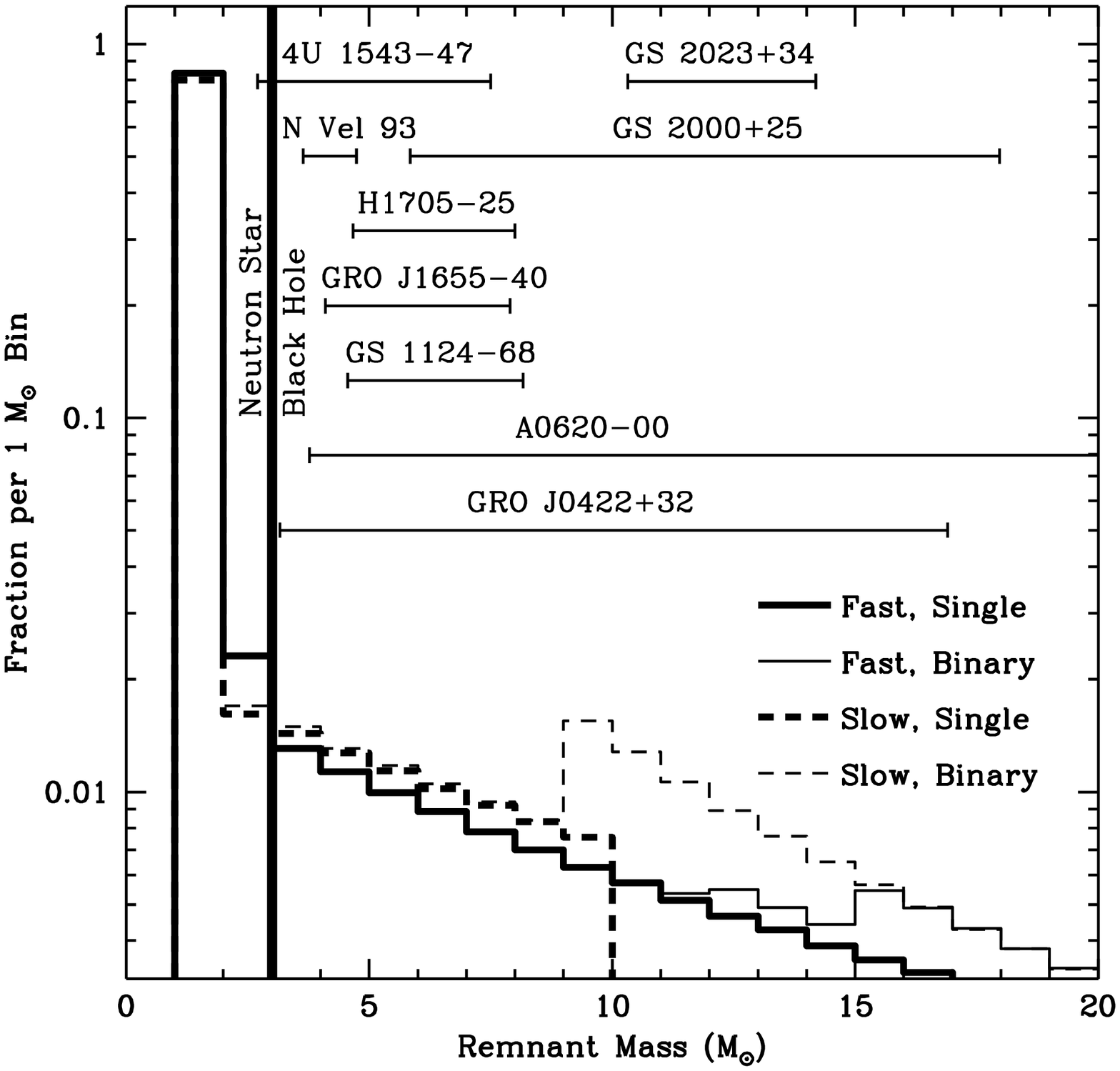}
 \caption{The mass distribution of compact remnants using the
remnant--progenitor mass relation in Figure 5 and assuming
an IMF with $\gamma=2.7$. Line types are the same as in Figure 5.}
 \end{figure}

\subsubsection{Wind Mass Loss from Single Stars}

The physical processes driving winds for hydrogen stars and Wolf-Rayet
stars are not entirely known and may be different for these two phases of
mass loss.  To study the effects of mass loss, we use the stellar models
of \cite{LH95}(1995) and \cite{S92}(1992), who adopt different
prescriptions for mass loss from hydrogen-rich and Wolf-Rayet stars.

\begin{figure}[ht]
 \epsscale{0.80}
 \plotone{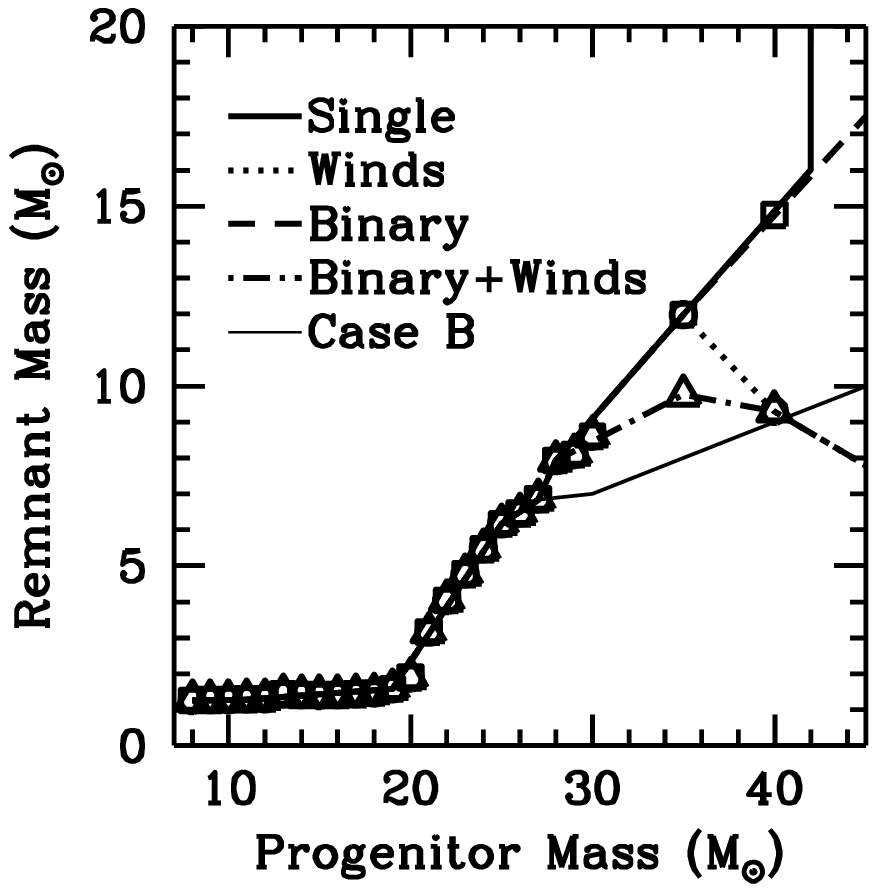}
 \caption{Remnant mass as a function of progenitor mass assuming 50\% of
the explosion energy from Figure 1 goes into unbinding the star.  The
different curves correspond to: single stars without winds (solid line),
single stars with winds (dotted line), binaries without winds (dashed
line), binaries with winds (dot-dashed line), and Case B (thin solid
line). }
 \end{figure}

Single stars less massive than $\sim$10\,M\sun\, are not affected by
winds. Stars with masses up to $\sim$40\,M\sun\, lose a significant
fraction of their hydrogen-rich envelope, most of it during the phase of
core helium burning, but their cores remain largely unaffected by mass
loss. Hence, the distribution of remnant masses from stars initially less
massive than $\sim 40$\,M\sun\, are not sensitive to winds from single
stars.  More massive BH progenitors, however, experience such a dramatic
decrease in their mass that they lose their entire hydrogen-rich envelope
before they reach collapse.  Further mass loss occurs during the strong
Wolf-Rayet phase of the exposed helium cores.  Wolf-Rayet winds considered
in the models are strong enough to remove much of the helium core, causing
most massive stars to form low-mass black holes (or even neutron stars).
The remnant mass actually decreases as a function of initial progenitor
mass (Figure 7, dotted line).  However, since these massive progenitors
are so rare, the primary effect of single-star winds on the BH mass
distribution is to limit the maximum BH mass to $\sim$12\,M\sun\, 
(Figures 7 and 8) and to make the slope of the distribution roughly flat for
remnant masses higher than 7\,M\sun\, up to the maximum (Figure 8) . We
note that for massive stars of lower metallicity, the wind loss rates are
lower and therefore the maximum BH mass is higher.

\begin{figure}[ht]
 \epsscale{0.80}
 \plotone{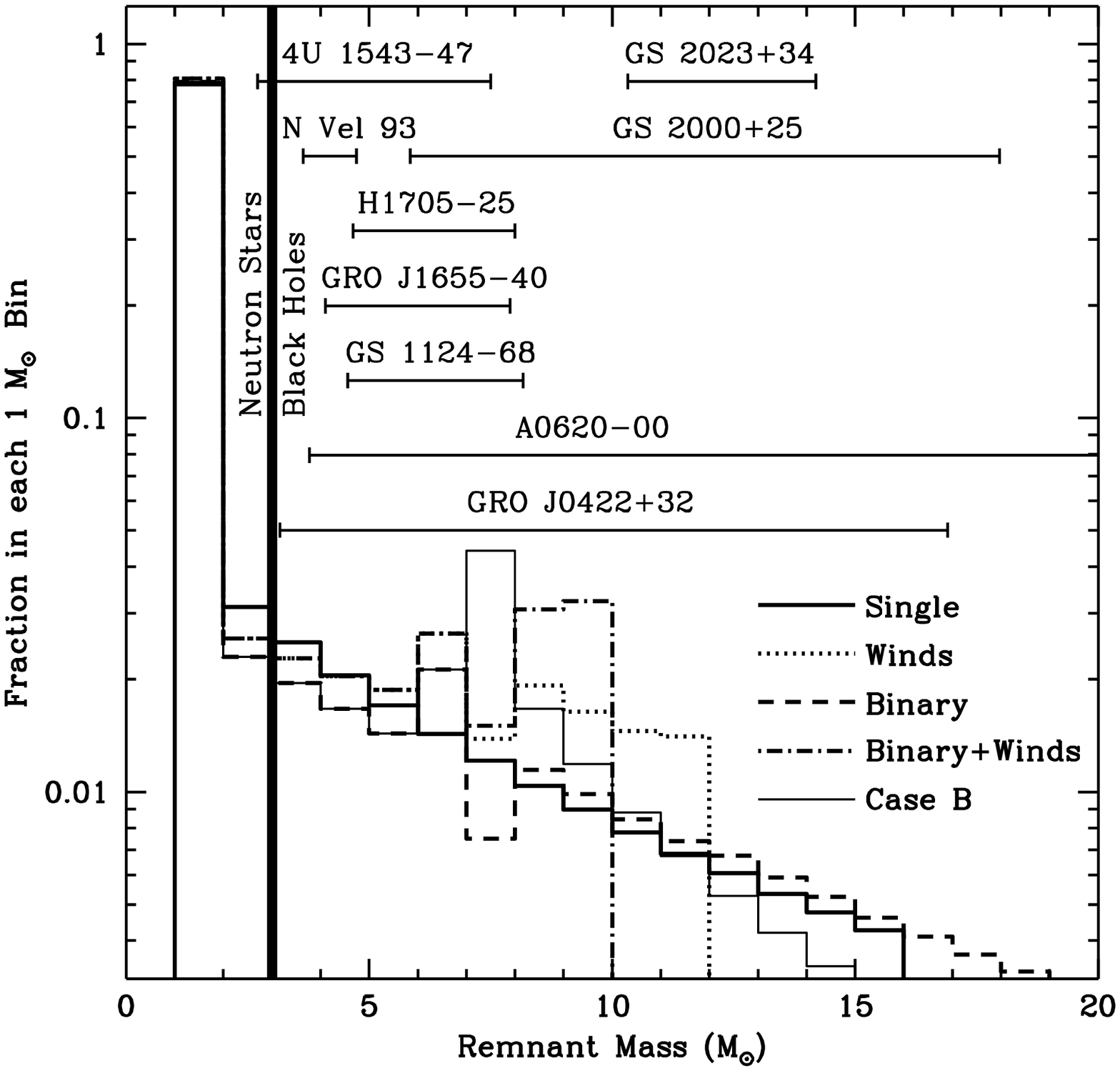}
\caption{The mass distribution of compact remnants using the
remnant--progenitor mass relation in Figure 7 and assuming
an IMF with $\gamma=2.7$. Line types are the same as in Figure 7.}
\end{figure}

\subsubsection{Winds and Black Hole Progenitors in Binaries}

So far, our discussion has been limited to single star evolution.  
However, BH masses are determined only in binaries.  The short orbital
periods of BH X-ray binaries ($<1$\,day to a few days, see \cite{C98}1998)
imply that their formation involves a common envelope phase
(\cite{K99a}1999a). Therefore, to compare with the observations we must
consider the effect of these {\it close} binaries on the BH mass
distribution.  A common envelope phase occurs when the BH progenitor star
expands and transfers mass onto its companion.  In many cases, the
companion is engulfed by the BH progenitor and it spirals into the
hydrogen envelope of the massive star, ejecting the hydrogen layers as it
spirals inward. As a result, the mass of the BH progenitor at collapse
decreases and consequently limits the BH remnant mass.  Mass ejection
because of binary evolution (e.g., common envelope) facilitates mass loss
from winds by uncovering the helium core and causing even those BH
progenitors with masses less than 40\,M\sun\, to enter a high mass-loss,
Wolf-Rayet phase.  The importance of binary effects on the BH mass
distribution depends upon how effective this Wolf-Rayet phase is at
removing mass, which in turn, depends upon when the common envelope phase
occurs (\cite{WLW95}1995).  As we shall show in this section, this depends
upon the separation of these binaries and our assumptions about the
evolution of stellar radii.

In the {\em absence} of Wolf-Rayet winds, binary evolution merely limits
the formation of very massive black holes (Figs. 7, 8). The hydrogen
layers of the BH progenitor are ejected during common-envelope evolution,
setting the mass of the star at collapse (and hence the maximum BH mass)
equal to the mass of its helium core.  If this occurs early in the
progenitor's life, before the star goes through hydrogen shell burning,
the helium-core masses tend to be somewhat smaller (by $\sim$1\,M\sun).  
However, this does not affect the inner core dramatically, and the
explosion and binding energies of these binary BH progenitors are roughly
the same as the values of single stars.  The primary effect of binaries
without winds is to constrain the black hole remnants of massive stars
which undergo direct collapse (no explosion) to the mass of the helium
core.  This produces more black holes with masses between 10--20\,M\sun,
but
does not allow many black holes to be more massive than 20\,M\sun.

Our conclusion that in the absence of helium-star winds, the remnant masses do 
not depend on the presence or absence of a hydrogen envelope seems to 
contradict what has been stated in the past, for example by 
\cite{W88}(1988), \cite{HW94}(1994), and \cite{F99a}(1999a). In these studies 
it was assumed that fallback was driven by the deceleration of the explosion 
shock as it enters the hydrogen envelope.  The most recent simulations of 
fallback (\cite{MF99}1999), indicate that the hydrogen envelope does not 
influence the fallback process. Therefore we do not expect the hydrogen 
envelope to affect the amount of fallback except in the most massive
stars, which in its presence would leave more massive remnants.

Because binaries facilitate the strong Wolf-Rayet winds by uncovering the
helium cores of moderate-mass BH progenitors, winds are much more
effective in binaries.  As we discussed in \S 2.2.2, Wolf-Rayet winds
significantly reduce remnant masses.  For single stars, these winds only
affect the very massive stars (and the very massive black hole remnants).  
However, in binaries, even stars less massive than 40\,M\sun\, experience a
Wolf-Rayet wind phase and this can affect the entire black hole mass
distribution. The amount of mass lost depends upon when the BH progenitor
loses its hydrogen envelope, that is, when mass transfer occurs.  If it
occurs before helium ignition (as the star expands off the main sequence),
the star will lose mass during its entire helium burning phase.  If mass
transfer does not occur before helium ignition, it will not occur until
after helium exhaustion because the star does not expand during helium
burning (\cite{S92}1992). After helium exhaustion, the star evolves so
quickly that even the strong Wolf-Rayet winds currently assumed in the
stellar models do not effect the mass significantly. We therefore have to
consider separately BH progenitors that lose their hydrogen envelopes
before (Case B mass transfer) or after (Case C mass transfer) core helium
ignition.

With the current Wolf-Rayet mass-loss rates and models for helium star 
evolution, those BH progenitors which undergo Case B mass transfer will 
not form black holes.  In this case, the exposed helium stars lose so much 
mass during the core helium burning that the BH progenitor's mass at 
collapse has decreased down to $\sim 3-4$\,M\sun\, (\cite{W95}1995; 
\cite{W99}1999). Such low-mass helium stars collapse to NS and not BH, 
despite the fact that their initial progenitor mass exceeds $\sim$20\,M\sun.

If black holes cannot be formed in any massive binaries which undergo Case
B mass transfer, the rate of BH formation decreases drastically
(\cite{PZ97}1997).  To avoid Case B mass transfer, the orbit must be
sufficiently wide to avoid mass transfer before helium ignition.  But
during core helium burning, winds cause the binary separation to widen
even further, so the BH progenitor must expand significantly to undergo
Case C mass transfer\footnote{Recall that mass transfer must occur to
tighten the binary orbit and produce the observed BH X-ray binaries.}.  
However, in typical stellar models of massive stars (e.g. \cite{S92}1992),
such expansion late in the evolution of the star occurs only for stars
with initial masses below $\sim$25\,M\sun\, (see also \cite{KW98}1998).  
This produces a narrow range from 20 to 25\,M\sun\, in which black holes
could actually formed.  Among binaries with primaries in the range
$20-25$\,M\sun, only a small fraction have orbits wide enough to avoid
Case B mass transfer but also tight enough to go through Case C evolution.  
This additional constraint is satisfied typically for orbital separations
in the range $1600-1800$\,R\sun\, (\cite{KW98}1998).  The fraction of BH
progenitors in binaries which satisfy all of these constraints is less
than 1\%!  (Table 2, Row 8) In addition to this drastic reduction in the 
BH X-ray binary formation rate, the current stellar models face an even more 
serious problem:  the black holes formed through this path are all less massive
than $\sim 7$\,M\sun\, (helium core mass of a 25\,M\sun\, star). Such an
upper limit on the BH mass cannot account for systems like V404 Cyg (GS
2023+34) with its BH mass in the range $10-14$\,M\sun.

Clearly something is wrong with the current stellar models.  Either the
current mass-loss rates from winds are too high or the radial evolution of
stars is incorrect (\cite{K99b}1999b).  We calculate BH mass distributions
for two different possible cases: (i) the wind models for helium stars are
correct but hydrogen rich stars more massive than 25\,M\sun\, experience
enough radial expansion after core helium exhaustion to evolve through
Case C mass transfer, and (ii) the models for the late evolution of
mass-losing massive stars are correct but wind mass loss for helium stars
significantly weaker than assumed so far. Then helium stars exposed
through Case B evolution lose smaller amounts of mass during core helium
burning and at collapse are massive enough to form black holes.

If we assume that Wolf-Rayet wind models are correct, but that the radial
evolution of stars are such that most stars undergo Case C mass transfer,
we can explain the currently observed black hole binaries.  In Case C mass
transfer, the helium core is not uncovered until after helium exhaustion
and the core collapses before any significant amount of mass is lost
through a Wolf-Rayet wind (\cite{W99} 1999).  Because these stars are in
binaries, the maximum black hole mass is limited to the helium core mass.  
In addition, because we assume that the current models for winds are
correct, stars more massive than 40\,M\sun\, still lose much of their mass
in winds, placing a further cap on the maximum black hole mass
(Binary+Winds curve in Figs.\ 7, 8).  With our choice in this case to
ignore the current calculations for the radial evolution massive stars, we
obtain a BH mass distribution where black holes make up $\sim$15\% of all
compact remnants (Table 2) and the maximum BH mass of 10\,M\sun\, (just
barely consistent with V404 Cyg).

Alternatively, it is possible that the true mass loss rates for helium
stars are lower than what is estimated by \cite{L89}(1989) and used by
\cite{WLW95}(1995) amd \cite{W99}(1999). In fact, empirical estimates of
Wolf-Rayet mass loss rates (\cite{HK98}1998) have decreased with time. To
investigate the effect of low Wolf-Rayet mass loss rates we calculate the
BH mass distribution assuming that (i) 99\% of the close binaries go
through Case B evolution (and the remaining 1\% go through Case C), and
(ii) the amount of mass lost from the helium star by the time it reaches
collapse is half of that calculated by \cite{WLW95}(1995).  This roughly
corresponds to losing at most half of the initial helium-star mass (see
\cite{K99a}1999a for constraints on the total amount of wind mass loss from
helium stars). The corresponding mass distribution (Case B in Figs.\ 7, 8)
produces black holes with masses in excess of 10\,M\sun\, and can easily
explain massive black holes such as V404 Cyg. However, this case may
require a drastic reduction in the mass loss rate to cut the total mass
lost by winds in half.  Mass loss rates depend sensitively upon stellar
luminosity and they decrease as the star loses mass.  Decreasing the mass
loss rate ($\dot{M}_{\rm wind}$) allows the star to retain its mass longer
(and its high luminosity) and hence, it stays in a rapid mass-losing phase
($t_{\rm wind}$) longer.  Since the total mass lost by winds is
$\dot{M}_{\rm wind}\times t_{\rm wind}$, lowering the mass loss rate need
not significantly lower the total mass lost (\cite{W99} 1999).

\section{Discussion}

In Table 2 we summarize our results for the range of theoretical BH mass
distributions in terms of the percentage of BH among collapse remnants in
different mass ranges. Despite the various uncertainties, some properties
of the distributions are quite robust. In all examined cases we obtain a
{\em continuous} remnant mass distribution that covers a wide range of
values, from NS up to BH masses of at least 10--15\,M\sun. The form of the
distribution is typically well described by an exponential of varying
steepness, which in some cases is accompanied by an upper mass cut-off. We
found no cases that lead to the formation of any kind of gaps in the
distribution, although the appearance of valleys in the distribution in
the range of $3-7$\,M\sun\, is possible (caused by winds and binaries -
Fig.\ 8).  The fraction of BH masses in the $3-5$\,M\sun\ range, varies from
15\% to 30\% and cannot be characterized as a gap. This fraction is mostly
sensitive to the assumed power-law index $\gamma$ of the massive-star IMF.
The fraction of remnants in the $5-10$\,M\sun\, range is more sensitive to
the assumptions about binaries and winds. It varies from 26-35\%, 46-57\%,
69-74\% for single stars, single stars with winds, and binaries with
winds, respectively.  Finally the fraction of remnants in the
$10-15$\,M\sun\, range is most sensitive to the assumptions about stellar
winds. These assumptions also determine the existence and position of an
upper BH mass cut-off.

Investigations of the observed sample of compact objects (\cite{B98}1998;
\cite{C98}1998)  conclude that there is a discontinuous jump from NS to BH
masses, that is, there exists a gap in the remnant mass distribution in
the range $3-5$\,M\sun.  \cite{B98}(1998) further found that the data
(excluding V404 Cyg) appear to be consistent with a narrow BH mass
distribution positioned at $\sim 7$\,M\sun. Our results appear to be in
disagreement with the conclusions in these studies. In what follows we
address this disagreement in the light of a number of biases that may be
operating and we derive some constraints on our theoretical limits.

There are three possible ways to resolve the apparent conflict between
models and observations: i) the observed sample used by \cite{B98}(1998)
and \cite{C98}(1998) suffers from low-number statistics and larger samples
will fill in the apparent gap at low BH masses and will best fit with
broader distributions, (ii) there are biases operating against the
identification or formation of X-ray binaries with low-mass BH, or (iii)
one or more parts of the theoretical calculation presented here is
incorrect.

Table 1 (and all the plots of our theoretical BH mass distributions)  
show the latest data on the mass estimates of the observed BH systems.
Note that for 7 out of the 9 systems the measured BH mass ranges extend to
values below 5\,M\sun. It is also important to point out that the current
sample includes two very strong candidates for low-mass BH (Nova Velorum
1993 and 4U1543-47). These two systems have been observed only recently
and were not included in the sample considered by \cite{B98}(1998). In
addition, new estimates of the mass of GRO J1655-40 predict a much lower
limit for the black hole mass (\cite{Phi99}1999). This result drastically
changes the statistical analysis of \cite{B98}(1998), which were dominated
by the small error bars of GRO J1655-40 around a mass of $\sim 7$\,M\sun.
Taking all the above into account it is possible that the conclusions
drawn from an earlier sample may not hold any longer.

\begin{figure}[ht]
 \epsscale{0.80}
 \plotone{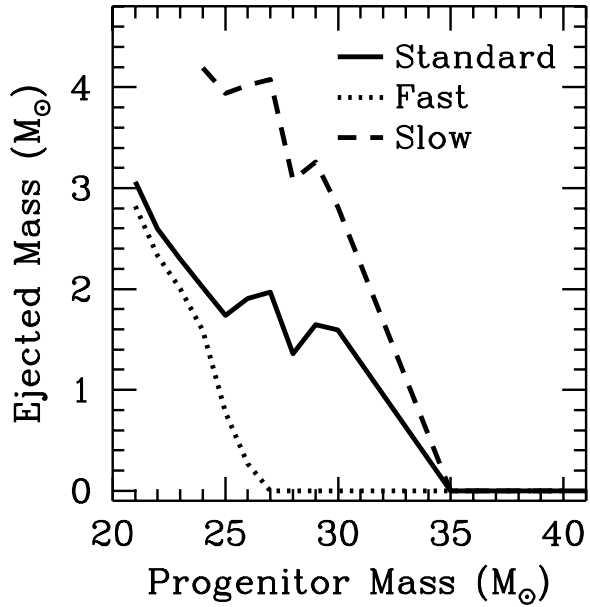}
\caption{Mass ejected during BH formation as a function of 
progenitor mass for three different cases:  our standard (SD) case from
Figure 1 (solid line) and the cases of fast (dotted line) and
slow (dashed line) rise/decline of the unbinding energy from Figure 4. We
assume that 50\% of the explosion energy is available to unbind the
envelope and that both binary and wind effects (B+W) are taken into
account.}
\end{figure}  

Another possibility is that there is some observational selection effect
against the identification of systems with low-mass black holes. The
transient character is one example. Dynamical measurements of masses of the
accreting objects in the systems studied so far can be made only during
their quiescent phase so the source must be transient. Systems with
lower-mass black holes will typically have mass ratios closer to unity
(given the low mass companions, $<1$\,M\sun, observed) than the detected
binaries. This will lead to higher mass transfer rates from the companions
and it is reasonable to expect that these systems are actually persistent
X-ray sources (and hence, mass measurements are hard to obtain). Such a
trend has been seen in the results obtained by \cite{KK97}(1997)  who
studied the transient character of BH binaries. Systems with an unevolved
low-mass companion ($\sim$1\,M\sun) move from unstable to the stable
regime as the mass of the BH decreases below $5$\,M\sun\, down to
$2$\,M\sun .

In the context of biases against these systems, it is also possible that
such systems have lower formation probability compared to higher-mass BH
systems because of events in their formation that are not considered here,
such as the effect of the BH formation (mass loss) on the binary
characteristics. Using our results for the remnant mass as a function of
progenitor mass (Figures 5, 7) we calculate the amount of mass lost during
BH formation as a function of progenitor mass (Figure 9). It is evident
that the formation of lower mass BH ($3-5$\,M\sun) is associated with a higher
fractional (relative to the progenitor or the BH) mass loss than the formation
of $\sim 7$\,M\sun\, black holes. Although this mass loss is not enough to
unbind the post-collapse system, even for the low-mass BH, it is enough to
affect the size of the orbits after BH formation. We find that, the degree
of orbital expansion immediately after the collapse in the case of
3\,M\sun\, BH is higher than for 7\,M\sun\, BH by a factor of $\simeq 3$.
So binaries with low-mass BH tend to be wider than binaries with higher
mass BH. Given the typical upper limit ($\sim$10\,R\sun) on orbital
separation for systems to reach Roche lobe overflow (and hence enter an
X-ray phase) (see \cite{K99a}1999a) and assuming a flat logarithmic
distribution of separations, we estimate that the number of X-ray
transients with 3\,M\sun\, BH (and low-mass companions) is lower than that
of 7\,M\sun\, BH by $\sim 35$\%. We find that this fraction is sensitive
to the form of the explosion energy dependence on progenitor mass
(standard, slow or fast).  However, this imbalance may not be high
enough to account for a possible gap.

\begin{figure}[ht]
 \epsscale{0.80}
 \plotone{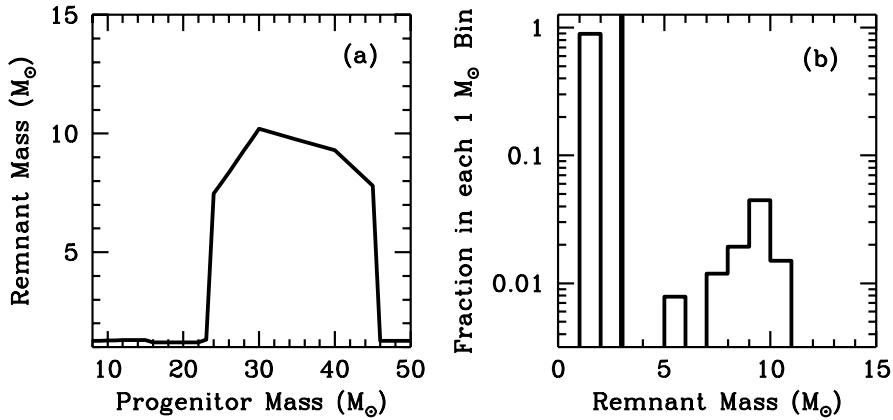}
\caption{(a) Remnant mass as a function of progenitor mass and (b)
distribution of remnant masses, assuming a step function for the
dependence of explosion energy on progenitor mass. A gap in the range of
2--5\,M\sun\, remnant masses appears (see text).}
\end{figure} 

If the number of observed BH transients increases sufficiently to overcome
the low-number statistics and observational biases, it is possible that it
will support the existence of a gap in the range $3-5$\,M\sun . Such a gap
then could place strong constraints on the supernova explosion energy and
its dependence on progenitor mass. One way a gap could be produced in the
appropriate range of remnant masses is if the explosion energy behaves as
a step function: e.g., $E_{\rm exp}=2.5\times10^{51}$\,ergs for $M_{\rm
prog}<$\,23\,M\sun\, and $E_{\rm exp}=0$\,ergs for $M_{\rm
prog}>$\,23\,M\sun).  For this distribution of supernova energies, a gap
in the remnant masses in the range 2--5\,M\sun\, appears (Figure 10).
However, the observed range in supernova energies and our current
understanding of the supernova mechanism both argue against supernova
energies which follow a discrete step function in their dependence on
progenitor mass, but a BH mass gap may force us to re-evaluate the current
picture.

\subsection{Neutron Star Mass Distribution}

Although our focus in this paper is the BH mass distribution, it is still
useful to examine our models with respect to what is known about NS
masses. The NS mass distribution is more sensitive to core-collapse models
and these results are somewhat more tentative than our BH mass
distribution.  As in the case of BH systems, a statistical examination of
the measured NS masses shows that they are consistent with a very narrow
distribution around 1.35\,M\sun\, (\cite{T99}1999). Note that the sample
used in that study does not include recent results on Vela X-1 with a
2-sigma lower limit on its NS mass of $\sim$1.6\,M\sun\, (\cite{B97}1997;
van Kerkwijk 1999, private communication). This narrow distribution agrees
well with our predicted mass distribution (Figure 11).  In our models,
81-96\% of neutron stars lie in the mass range between 1.2--1.6\,M\sun.
Beyond 1.6\,M\sun, there is a continuous distribution of neutron stars out
to the maximum neutron star mass. The results are quite insensitive to the
various uncertainties (explosion energy behavior and fraction available to
unbind the envelope, binary and wind effects). \footnote{However, the initial 
mass of the proto-neutron star (without fallback) is extremely sensitive 
to the supernova explosion mechanism which is not completely understood.}. 
We would expect 1 or 2
neutron stars with masses beyond 1.6\,M\sun in the current data, and Vela
X-1 seems to fall in this range.  Our distribution is to be contrasted
with that derived by \cite{T96}(1996), which has a double peaked neutron
star distribution.  Their distribution may also fit the observed sample of
neutron stars with one peak explaining the cluster of neutron stars at
1.35\,M\sun, and the other fitting neutron stars such as Vela X-1.
However, those results were obtained without considering the effects of
fallback, which would smooth out the double-peaked distribution that was
reported.
\begin{figure}[ht]
 \epsscale{0.80}
 \plotone{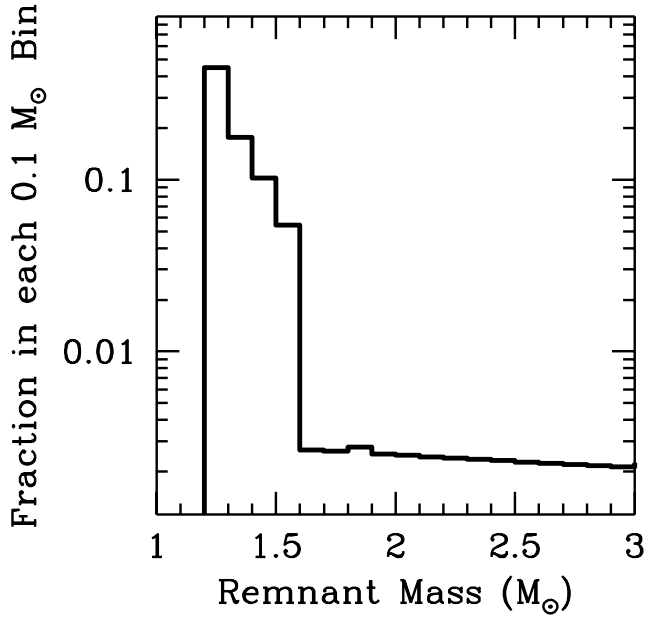}
\caption{Distribution of neutron star remnant masses for 
our standard case for the explosion energy from Figure 1 ($f=0.5$).  }
\end{figure}

\subsection{Effects of Winds and Binary Evolution}

Our study demonstrates the strong sensitivity of the formation of massive
black holes ($>10$\,M\sun) to the effects of winds.  In close binaries
(which form the observed black hole systems) these effects become even
more extreme.  In the simple case of single stars, winds alter the shape
of the BH mass distribution by making it flatter and decreasing the
maximum BH mass to $10-15$\,M\sun. When the effect of a close companion is
taken into account, winds drastically reduce the formation rate of BH and
cause the maximum BH mass to decrease even further (at or below
10\,M\sun).  Combined with the current stellar models (and in particular,
the radial evolution of stars), theory can not explain the observed black
hole systems (\cite{K99b}1999b).  The resolution of these problems lies
either in significantly lowering mass loss rates, so that helium stars
lose less than half of their initial mass (see also \cite{K99a}1999a), or
in widening the range of both progenitor masses {\em and} binary orbital
separations over which evolution through Case C mass transfer is allowed.
The latter of the two possibilities would require that current estimates
of stellar radii at the end of core helium burning in massive stars are
actually {\em over}estimates {\em and} that the stellar radii just prior
to collapse are actually {\em under}estimates, especially for stars more
massive than $\sim 25$\,M\sun.  Given the uncertainties involved in
theoretically determining the radii of evolved stars, such modifications
are still possible.  If revised stellar radii can not explain the black
hole mass distribution, then the observed black hole masses require that
mass loss from stellar winds must decrease.  The downward trend of
mass-loss rates from Wolf-Rayet winds (\cite{HK98}1998) suggests that this
may be the correct solution.

\subsection{Neutron Star and Black Hole Kicks}

The evidence that neutron stars receive kicks during their formation
continues to grow (see \cite{FWH99}1999 for a review)  and it is generally
accepted that neutron stars must receive kicks with a mean velocity
exceeding $\sim$100-200\,km\,s$^{-1}$ up to $\sim 400$\,km\,s$^{-1}$.  The
question of BH kicks is still open however.  Black hole X-ray binaries
seem to have a Galactic distribution with a scale height smaller than that
of NS X-ray binaries and an inferred mean velocity of only $\sim
40$\,km\,s$^{-1}$ (\cite{WvP96}1996).  \cite{TC97}(1997) and
\cite{NT99}(1999) have argued that these velocities can be explained
simply by mass ejection during BH formation and hence kicks are not
necessary to explain the properties of BH binaries.

In constrast the majority of BH X-ray transients, GRO J1655-40 appears to
have a large space velocity.  \cite{B95}(1995) measured its radial
velocity to be 114$\pm19$\,km\,s$^{-1}$, which provides a lower limit to
the space velocity. Under the assumption of isotropicity the measured
radial velocity corresponds to a 3-D velocity of $\sim 200$\,km\,s$^{-1}$.
\cite{Phi99}(1999) were able to constrain the magnitude of the space
velocity in the range $143-153$\,km\,s$^{-1}$. \cite{NT99}(1999) used the
current determinations of the BH mass and a relatively high mass for the
BH companion and obtained space velocities of at most 110\,km\,s$^{-1}$
(just compatible with the radial velocity measurement). In their study the
amount of mass loss was treated as a free parameter. However, as we
discussed earlier in the paper, our calculations link the BH mass to the
amount of mass lost during BH formation. We use our results for the full
set of cases (different dependence of explosion energy on the progenitor
mass and different values of the parameter $f$) and find that space
velocities in the range $100-200$\,km\,s$^{-1}$ require pre-collapse
orbital separations ($0.1-4$\,R\sun) too small to accommodate the BH
companion in the binary. Imposing this latter constraint, we obtain
maximum space velocities, for the different models, in the range
$20-85$\,km\,s$^{-1}$. We conclude that the measured radial velocity of
GRO J1655-40 seems to require that a kick was imparted to the BH (in
agreement also with \cite{BPS95}1995).  Given that the majority of the
systems appear to have low space velocities (\cite{WvP96}1996),
\cite{F99b}(1999b) found that a BH kick of at least $\sim
50$\,km\,s$^{-1}$ is required to explain the velocities of the complete
set of BH X-ray transients.

We have shown here that most of the black holes are formed via fallback
and not promptly. Since many of the NS kick mechanisms operate before
fallback occurs, it seems reasonable to expect that BH receive kicks as
well.  Given the kick mechanisms, it is likely that the momentum imparted
by asymmetric explosions is equal for both BH and NS, and therefore the
kick scales with mass:  $<V_{\rm BH}>=<V_{\rm NS}> M_{\rm NS}/ M_{\rm
BH}$.  If this is the case, then lower mass BH will receive higher kicks.  
This, in the context of a possible gap in the low-mass BH range, possibly
provides an additional bias against the formation of low-mass BH in
binaries. We find that, for a typical NS kick magnitude of
200\,km\,s$^{-1}$ and for reasonable values of post-CE orbital
separations, 25\%--50\% (for a fixed BH kick magnitude of 50\,km\,s$^{-1}$
the fraction is in the range 10\%-50\%) of the systems with 3\,M\sun\, BH
get disrupted while systems with 7\,M\sun\, BH remain unaffected. This
effect could contribute to decreasing the formation frequency of X-ray
binaries with low-mass black holes. 

\acknowledgements We thank N.\ Langer, S.\ Wellstein, and A.\ Heger 
for useful discussions.  The work of CLF was supported by 
NASA (NAG5-8128) and the US DOE ASCI Program (W-7405-ENG-48).  
VK acknowledges support by the Smithsonian Institution in the form 
of a CfA Postdoctoral Fellowship.

\begin{deluxetable}{lc}
\tablewidth{30pc}
\tablecaption{Measured Black Hole Masses\tablenotemark{a}}
\tablehead{\colhead{System} & \colhead{Mass Range (M\sun)}}

\startdata

 GS 2023+34 (Nova Cyg 1938/1989) (V404 Cyg) & 10.3-14.2\tablenotemark{b}
\nl
 GS 2000+25 (Nova Vul 1988) & 5.84-18.0 \nl
 H1705-25 (Nova Oph 1977) & 4.67-8.00 \nl
 GRO J1655-40 (Nova Sco 1994) & 4.1-7.9\tablenotemark{c} \nl
 GS 1124-68 (Nova Mus 1991) & 4.56-8.17 \nl
 A0620-00 (Nova Mon 1975) (V616 Mon) & 3.78-25.4 \nl
 GRO J0422+32 (Nova Per 1992) & 3.16-16.9\tablenotemark{d} \nl
 GRS 1009-45 (Nova Vel 1993) (V1234 Oph) &
$\sim$3.64-4.74\tablenotemark{e} \nl
 4U1543-47 & $\sim$2.7-7.5\tablenotemark{f} \nl

 \tablenotetext{a}{We do not include the high mass X-ray binaries (Cyg
X-1, LMC X-1, LMC X-3), whose masses are much less certain than those
obtained in X-ray transients.}
 \tablenotetext{b}{Unless otherwise stated, data are taken from 
\cite{B98}(1998).}
 \tablenotetext{c}{The lower limit for the black hole mass comes 
from the analysis of \cite{Phi99}(1999).  The upper limit comes 
from the revised mass function of \cite{Sh99}(1999).}
 \tablenotetext{d}{We have included the revised mass ratio and mass
function from \cite{Har99}(1999).  Note that \cite{Bee97}(1997) have
argued that the BH mass in this system may be much larger than our upper
limit ($^>_\sim$28M\sun).  They argue that the inclination angle is much
lower than past estimates.}
 \tablenotetext{e}{This system is not yet studied in detail
(\cite{Fil99}1999) and the error bars may be revised in the
future.}  
 \tablenotetext{f}{Like Nova Vel 1993, this system is not yet studied in
detail (\cite{O98}1998).}

\enddata
\end{deluxetable}
\clearpage

\begin{deluxetable}{lccccccc}
\tablewidth{36pc}
\tablecaption{Black Hole Masses}
\tablehead{ \multicolumn{3}{c}{Model Parameters} &
\multicolumn{5}{c}{Percentage of BH Compact Remnants
$\equiv N_{\rm BH}/(N_{\rm BH}+N_{\rm NS})$} \\
\colhead{Energy\tablenotemark{a}} & \colhead{Winds\tablenotemark{b}}
& \colhead{IMF\tablenotemark{c}} & \colhead{Total} &
\colhead{3-5M\sun} & \colhead{5-10M\sun} & \colhead{10-15M\sun} &
\colhead{$>$15M\sun}}

\startdata

SD, 100\% & ----- & 2.0 & 29.9 & 5.0 & 8.2 & 4.7 & 12 \\
SD, 100\% & ----- & 3.0 & 12.1 & 3.1 & 4.1 & 1.9 & 3.0 \\
SD, 10\% & ----- & 2.0 & 42.1 & 8.1 & 12.2 & 6.8 & 15.0 \\
SD, 10\% & ----- & 3.0 & 21.3 & 6.5 & 7.6 & 3.1 & 4.1 \\
SD, 50\% & ----- & 2.7 & 19.1 & 4.6 & 6.3 & 3.1 & 5.1 \\
SD, 50\% & B & 2.7 & 18.5 & 3.6 & 6.4 & 3.4 & 5.1 \\
SD, 50\% & W & 2.7 & 16.7 & 4.3 & 9.5 & 2.9 & 0 \\
SD, 50\% & B+W$^*$ & 2.7 & 0.088 & 0.043 & 0.045 & 0 & 0 \\
SD, 50\% & B+W & 2.7 & 16.6 & 4.3 & 12.3 & 0 & 0 \\
SD, 50\% & Case B & 2.7 & 18.5 & 3.6 & 10.4 & 2.9 & 1.5 \\
Slow, 50\% & ----- & 2.7 & 14.2 & 2.4 & 4.0 & 2.4 & 5.4 \\
Fast, 50\% & ----- & 2.7 & 18.4 & 2.7 & 4.7 & 0.30 & 10.7 \\
Slow, 50\% & B & 2.7 & 14.2 & 2.4 & 4.0 & 2.6 & 5.2 \\
Fast, 50\% & B & 2.7 & 18.2 & 2.8 & 5.6 & 4.6 & 5.2 \\
Slow, 50\% & W & 2.7 & 12.4 & 3.1 & 7.1 & 2.2 & 0. \\
Fast, 50\% & W & 2.7 & 16.5 & 3.4 & 7.6 & 2.7 & 2.8 \\
Slow, 50\% & B+W & 2.7 & 12.3 & 3.2 & 9.1 & 0. & 0. \\
Fast, 50\% & B+W & 2.7 & 16.4 & 3.6 & 11.3 & 1.5 & 0. \\

 \tablenotetext{a}{Explosion Energy:  SD$\equiv$standard rise/fall slope
-- see Figure 1, Slow, Fast$\equiv$slow, fast rise/fall slope -- see
Figure 4. 10,50,100\%: Fraction of explosion energy available to eject
material -- see Figure 1.}
 \tablenotetext{b}{Winds, Binary Effects: -----$\equiv$no winds or 
binary effects, B$\equiv$binary effects only, W$\equiv$Wind effects 
only, B+W$^*\equiv$ Winds+Binaries (stellar radii from literature), 
B+W$\equiv$Winds + Binaries (all binaries in Case C), Case
B$\equiv$Winds + Binaries (almost all binaries in Case B);  see text.}
 \tablenotetext{c}{Initial Mass Function power law $\gamma$, see text.} 

\enddata
\end{deluxetable}
\clearpage

\end{document}